\documentstyle[amssymb,12pt,thmsa,sw20lart]{article}

\input{tcilatex}

\begin{document}

\title{The Superconducting Proximity Effect as a Tool to Investigate Metal Films
and Interfaces.}
\author{D. Garrett, M. Zhang and G. Bergmann. \\
University of Southern California}
\date{\today }
\maketitle

\begin{abstract}
The superconducting proximity effect is measured in sandwiches of thin Pb
films and the alkali metals Cs, Rb, K and Na. The $T_{c}$-dependence
provides information about the interface barriers between Pb and the
alkalis. Such a barrier is particularly large in Pb/Cs sandwiches. It is not
due to impurities or oxydation. In the presence of a sufficiently strong
barrier a special form of the Cooper limit can be applied to calculate the
transition temperature of the sandwich.

PACS: 74.50.+r, 73.20.-r, 71.20.Dg

\newpage
\end{abstract}

\section{Introduction}

\medskip When a thin superconducting film is covered with a normal conductor
its transition temperature is lowered. This phenomenum is known as the
\textquotedblright Superconducting Proximity Effect\textquotedblright\
(SPE), and it was in tensively studied in the 1960s and 70s \cite{H25}, \cite
{H26}, \cite{W32}, \cite{D36}, \cite{H27}, \cite{B133}, \cite{M45} (see for
example \cite{D32}, ). (In those days one had the hope that an extrapolation
of $T_{c}$ would yield a finite transition temperature for normal conductors
such as the noble metals.) \ However, there has been a continuous interest
in this effect over the years \cite{Z7}, \cite{Z8}, \cite{Z9}, \cite{S39}, 
\cite{V11} which extended recently into superconductor-ferromagnetic metal
sandwiches (see for example \cite{S40}). In the present paper we want to
revive the SPE as a tool to investigate interfaces between metal films.

In recent years we have investigated the properties of thin alkali metal
films \cite{B111}, \cite{B121}, \cite{B122},\cite{B123}, \cite{B124}, \cite
{B129}, \cite{B130}, \cite{B131}, \cite{B132}. We observed a number of
properties which were quite unexpected:

\begin{itemize}
\item  The Hall effect and resistance of thin Cs increased dramatically when
the film was covered in situ with small concentrations of (s,p) impurities.

\item  Sandwiches of quench-condensed Fe covered with a film of Cs or K
showed a mean free path in the alkali film which can be up to five times the
thickness of the alkali film. This means that the electrons in the Cs or K
are almost perfectly specularly reflected at the free surface and the
interface with the Fe. The latter is quite surprizing since the Fe is very
disordered and the electrons should be diffusively scattered at the
interface.

\item  Recently we used the SPE in a Pb/K/Pb sandwich to study the local
electronic properties of the K while it was coveraged with sub-mono-layers
Pb. On the one hand the coverage of the K with Pb appeared to localize the
electrons in the K while on the other hand the $T_{c}$-reduction of the
first Pb film by the K remained unaltered.
\end{itemize}

\medskip The properties of interfaces and the ability of electrons to cross
them are important in many physical phenomena and applications. To name one
example for the latter, the giant magneto-resistance in magnetic
multi-layers depends critically on the ability of the conduction electrons
to cross ''freely'' from one film to the next. The goal of this paper is to
investigate the SPE in Pb/Ak films (in the following we use the symbol
''Ak'' as synonym for any of the four alkali metals Cs, Rb, K and Na) and to
show that the SPE is a suitable tool to obtain information about interfaces.

\section{Experiment}

In the present investigation we prepare sandwiches of Pb/Ak at liquid helium
temperatures. In most experiments we quench condense a Pb film of 13 to 14 $%
nm$ thickness and a resistance of about 100$\Omega $. The Pb is then covered
in several steps with an alkali metal. After each evaporation the film or
sandwich is annealed, the original Pb film up to 40 $K$ and the sandwiches
up to 35 $K$. Then the superconducting transition curve is recorded and the
magneto-resistance and Hall resistance are measured at 9.5 $K$ in the field
range between -7 $T$ and +7 $T$. In these experiments one has to be careful
that the edges of the Pb film are covered by the normal metal. Otherwise one
obtains double transition curves. Fig.1 shows a set of transition curves for
Pb/K sandwiches.

In Fig.2a-d the dependence of $T_{c}$ on the coverage with Cs, Rb, K and Na
is plotted. The full circles represent the experimentally measured $T_{c}$.
The curves are discussed below.

\section{Theory and Discussion}

Werthamer \cite{W32} derived an implicit set of equations for the transition
temperature of a sandwich of two superconductors. The superconductor with
the lower transition temperature is generally called the normal conductor
because at the $T_{c}$ of the sandwich it would be in the normal conducting
state. In this theory the gap function $\Delta \left( {\bf r}\right) $ in
the superconductor is proportional to $\cos \left[ k_{s}\left(
z+d_{s}\right) \right] $ and in the normal conductor proportional to $\cosh %
\left[ k_{n}\left( z-d_{n}\right) \right] $, with $z=0$ at the interface.
The $k_{s,n}$ are the inverse superconducting coherence lengths. For
disordered (dirty) metal films they are given by 
\begin{eqnarray}
\ln \left( \frac{T_{s}}{T_{c}}\right) &=&\chi \left( \xi
_{s}^{2}k_{s}^{2}\right) \text{, }\ln \left( \frac{T_{c}}{T_{n}}\right)
=-\chi \left( -\xi _{n}^{2}k_{n}^{2}\right)  \label{Wh1} \\
\chi \left( z\right) &=&\Psi \left( \frac{1}{2}+\frac{1}{2}z\right) -\Psi
\left( \frac{1}{2}\right)  \nonumber \\
\xi ^{2} &=&\frac{\hbar v_{F}l}{6\pi k_{B}T_{c}}=\frac{\pi \hbar k_{B}}{%
6e^{2}T_{c}}\frac{\sigma }{\gamma }=D\tau _{T}  \nonumber \\
\tau _{T} &=&\frac{\hbar }{2\pi k_{B}T_{c}}  \nonumber
\end{eqnarray}
Here $T_{s},$ $T_{n}$ and $T_{c}$ are the transition tempertures of the two
superconductors ($T_{s}>T_{n}$) and the sandwich $\xi _{s,n}$ are the
thermal coherence lengths, $\sigma $ the conductivity, $\gamma $ the
Sommerfeld constant (which stands for the density of states), $D$ the
diffusion constant, $v_{F}$ the Fermi velocity, $\tau _{T}$ the thermal
time, and $\psi \left( x\right) $ the digamma function.

At the interface Werthamer originally used the boundary condition that $%
\frac{1}{\Delta }\frac{d\Delta }{dz}$ is continous. de Gennes \cite{D12}
derived instead that $\Delta /\left( NV\right) $ and $\left( D/V\right)
\left( d\Delta /dz\right) $ are continous at the interface, the latter only
in the ''dirty'' limit. The combined function $\left( D/V\right) \left(
d\Delta /dz\right) /\Delta /\left( NV\right) =DN\left( d\Delta /dz\right)
/\Delta $ is then continous as well. This yields boundary condition 
\begin{equation}
N_{s}\xi _{s}^{2}k_{s}\tan \left( k_{s}d_{s}\right) =N_{n}\xi
_{n}^{2}k_{n}\tanh \left( k_{n}d_{n}\right)  \label{Wh2}
\end{equation}
which determine $T_{c}$. The equations (\ref{Wh1}) and (\ref{Wh2}) together
are often called the Wertheimer-deGennes theory. We will abriviate the
theory as the WG-theory.

The WG-theory is restricted to the dirty case, where the mean free path of
the conduction electrons is much smaller than the superconducting coherence
length. It is interesting to note that Werthamer's equations require only
three parameter of each superconductor besides the thickness: the transition
temperture, the resistivity and the density of states (Sommerfeld constant).

Deutscher and DeGennes \cite{D32} point out the calculation for $\Delta
\left( x\right) $ within each superconductor uses only one root for the
decay constant (the smallest Matsubara frequency). Nevertheless Werthamer's
equations were successfully applied to a number of experimental results (see
for example \cite{D32}, \cite{B133}). Therefore it is interesting to compare
the experimental results with the WG-theory.

In Fig.2a-d we calculate the transition temperature with the WG-theory for
the different Pb/Ak sandwiches. As far as we know the alkali metals are not
superconducting. On the other hand the WG-theory requires a finite
transition temperature $T_{n}$ for both metal films. Therefore we treat the
alkali metals as hypothetical superconductors with a transition tempereture
of $T_{n}=10^{-5}K$. (It turns out that the choice of $T_{n}=10^{-2}K$ does
not make any difference in the theoretical transition temperatures at the
experimental thicknesses of the alkali metals. Only in an extremely small
thickness range of the alkali metals $d_{n}\rightarrow 0$ does one find a
small difference (which, however, influences the initial slope $%
dT_{c}/dd_{n} $ at zero thickness of the alkali metal.) We recognize from
Fig.2 that the heavy alkali metals such as Cs show a large deviation from
the WG-theory while the sandwich with the light Na is relatively close to
the theoretical curve. For Pb/Rb and Pb/K the region of small alkali
thickness appears to be reasonably reproduced by the WT.

The initial slope $dT_{c}/dd_{n}$ of $T_{c}$ versus the thickness of the
normal metal at $d_{n}=0$ can be derived from Werthamer's theory and is
given by 
\[
\frac{d_{s}}{T_{s}}\frac{dT_{c}}{dd_{n}}=-\frac{\pi ^{2}}{4}\frac{N_{n}}{%
N_{s}}\chi ^{-1}\left( -\ln \left( \frac{T_{s}}{T_{n}}\right) \right) 
\]
The expression $\chi ^{-1}\left( -\ln \left( \frac{T_{s}}{T_{n}}\right)
\right) $ approaches the value one for a large $\left( T_{s}/T_{n}\right) $
ratio, but for $T_{n}=10^{-2}K$ (with $T_{s}=7.2K)$ one has $\chi
^{-1}\left( -\ln \left( 7.2/10^{-2}\right) \right) =-0.75$. Since the
Werthamer value of the initial slope depends strongly on the choice of $%
T_{n} $ while at finite thickness there is no such dependence we prefer to
compare the value $\Delta T_{c}/\Delta d_{n}$ between experiment and theory.
Here $\Delta d_{n}$ is the smallest evaporated film thickness of the alkali
metal and $\Delta T_{c}$ is the $T_{c}$-reduction due to this alkali
coverage. (We will still call the ratio $\Delta T_{c}/\Delta d_{n}$ the
initial slope.) These values are collected in Table I for the four alkali
sandwiches studied. For the alkali metals we used the density of states as
given in Ashcroft and Merman \cite{A30}. In collumn three the factor $%
N^{*}/N_{0},$ the ratio between the (experimental) density of states $N^{*}$
and the free electron value $N_{0}$, is given. For the very disordered Pb we
used a density ratio of 2.5 because disordered Pb has an even higher gap
ratio $2\Delta _{0}/k_{B}T_{c}=4.6$ than that of pure Pb and therefore a
larger value of $\left( 1+\lambda \right) $ than pure (annealed) Pb, where $%
\left( 1+\lambda \right) $ is the electron-phonon enhancement factor.

\[
\begin{array}{l}
\begin{tabular}{|l|l|l|l|l|l|l|}
\hline
$
\begin{array}{l}
{\bf Alkali} \\ 
\text{{\bf metal}}
\end{array}
$ & 
\begin{tabular}{l}
{\bf exper.} \\ 
{\bf code}
\end{tabular}
& $\frac{N^{*}}{N_{0}}$ & d$_{\text{n}}$ ($nm$) & $\left( \frac{\Delta T_{c}%
}{\Delta d_{n}}\right) _{\exp }$ & $\left( \frac{\Delta T_{c}}{\Delta d_{n}}%
\right) _{\text{Wh}}$ & $\frac{\tau _{T_{c}}}{\tau _{f}}$ \\ \hline
Na & GI & 1.3 & 2.18 & 0.280 & 0.32 & --- \\ \hline
K & GF & 1.2 & 2.04 & 0.235 & 0.27 & 0.08 \\ \hline
Rb & GC & 1.3 & 1.77 & 0.249 & 0.25 & 0.052 \\ \hline
Cs & OC & 1.5 & 2.29 & 0.175 & 0.31 & .035 \\ \hline
&  &  &  &  &  &  \\ \hline
\end{tabular}
\\ 
\text{Table I}:\text{The experimental value of }\frac{\Delta T_{c}}{\Delta
d_{n}}\text{ and the }\text{prediction according } \\ 
\text{to Werthamer's thoery }\text{for the different alkali sandwiches.}
\end{array}
\]

The initial slope does not depend on the mean free path in either metal. The
same is true for the small values of $\Delta d_{n}$. In Werthamer's theory
it only depends on the ratio of the density of states. For the Pb sandwiched
with Na, K and Rb the experimental initial slope agrees within 15\% with the
prediction of the WG-theory. For Cs the experimental initial slope is much
smaller than Werthamer's prediction. We believe that this is caused by some
kind of barrier at the Pb-Cs interface.

From the experimental results the deviation between experiment and WG-theory
for the Pb/Cs sandwich is particularly striking. One might object that the
small reduction in $T_{c}$ is caused by an oxide layer between the Pb and
the Cs. We exclude this interpretation for several reasons: (i) The results
have been reproduced in several experiments, (ii) the other alkali metal
such as Na and K would show an even stronger tendency to oxidation.

To investigate this question we prepare a sandwich in which the Pb is first
covered with 2 $nm$ of Na and afterwards with Cs of increasing thickness. In
Fig.3 the transition temperature of the Pb/Na/Cs sandwich is plotted versus
the alkali thickness $d_{Na}+d_{Cs}$. For comparison the $T_{c}$-dependence
of the Pb/Cs and the Pb/Na sandwiches are shown in the same figure. One
realizes that the transition temperature of the Pb/Na/Cs sandwich is much
closer to the Pb/Na sandwich than to the Pb/Cs sandwich. This demonstrates
that, after bridging the contact between the Pb and the Cs, one finds a
similar $T_{c}$-reduction as in the Pb/Na sandwich. It is not the electronic
properties of the Cs which yield the small initial slope and the weak
reduction of $T_{c}$ in the Pb/Cs sandwich. Instead the interface between Pb
and Cs must create some kind of obstacle which makes it harder for the
electrons to cross the interface. About the nature of this obstacles one can
only speculate at the present time. However, we are quite sure that it is
not due to dirt or oxygen. And neither the Pb nor the Cs are individually
responsible for the obstacle because in the Pb/Na/Cs sandwich the more
agressive Na faces each of the two and the obstacle (at each interface) is
either absent or much smaller than in the Pb/Cs interface.

If one compares the experimental $T_{s}$ curves for Pb with Cs, Rb, K and Na
with the theory of Wertheimer one recognizes right away that the deviation
between experiment and the Wertheimer theory decreases in going from Cs to
Na. It is not obvious that the Wertheimer theory has to be correct for our
Pb/Ak sandwiches; in fact this theory may be not appropriate at all for the
alkali metals. But the Pb/Na/Cs sandwich demonstrates for the Pb/Cs case
that the deviation is not due to the Cs but the interface. Therefore it is
suggestive that all the interfaces Pb/Ak present some kind of obstacles for
the transmission of the electrons through the interface, where the strength
of the obstacle decreases in going from Pb/Cs to Pb/Na. This does not
contradict the fact the experimental initial slopes for the three alkali
metals agree quite well with the Wertheimer theory. This we demonstrate for
the Cooper limit of the SPE.

The Cooper limit applies when the gap function $\Delta \left( {\bf r}\right) 
$ can be treated as constant in each metal of the film. This is fulfilled
when the thickness of each metal is much smaller than its coherence length.
If the exchange of electrons between the superconductor and the normal
conductor is strongly reduced then the Cooper limit is valid in a
considerably larger thickness range. As discussed in the Appendix A one can
formulate the following condition: Mark all electrons in a small energy
range about the Fermi surface in the superconductors. Follow their density
distribution as a function of time while they can propagate into the normal
conductor. If their density is at all times reasonably constant within the
superconductor, then the gap function $\Delta \left( {\bf r}\right) $ will
be a constant $\Delta _{s}$ (at the transition temperature) and the Cooper
case applies. This can, for example, happen for relatively thick films if
the escape time from the superconductor into the normal conductor is long.

In Appendix A we derive the Cooper limit with a barrier between the two
metals from a special version of the linear gap equation. The theory
requires one fitting parameter, the transmission rate $1/\tau _{sn}$ from
the Pb into the alkali film. The resulting theoretical curves are plotted in
Fig.2a,b,c as full curves and the fitted rates $\tau _{T_{c}}/\tau _{t}$ are
collected in table I. They describe the behavior of $T_{c}$ at finite
thickness $d_{n}$ of the alkali metal films quite well, in particular the
saturation in the Pb/Cs sandwich. Surprisingly the WG-theory gives a better
fit for the sandwiches with Rb, K and Na at very small alkali thicknesses.

For low transmission rate $1/\tau _{sn}$ through the interface and large
normal conductor thickness the electrons of the superconductor escape with a
rate of $1/\tau _{sn}$ from the superconductor. This acts as a
pair-weakening rate of $1/\tau _{sn}$ and reduces the transition temperature
as 
\[
\frac{\Delta T_{c}}{T_{s}}=\frac{\pi ^{2}}{2}\frac{\hbar }{2\pi k_{B}T_{c}}%
\frac{1}{\tau _{ns}} 
\]

Ashida et al. \cite{A57} calculated the transition of superconductor -
normal conductor sandwiches with barries inbetween. They described the
strength of the barrier by the coefficient of reflectivity $R$ at the
interface. (In the absence of a barrier the value of $R$ is not zero but
given by density of states ratio). They find in the limit $\left( R-1\right)
<<1$ for the transition temperature 
\[
\frac{\Delta T_{c}}{T_{s}}=\frac{\pi ^{2}}{16}\frac{\hbar }{2\pi k_{B}T_{c}}%
\frac{\left( 1-R\right) v_{s}}{d_{s}} 
\]
The two result are compared in the appendix.

A full numerical solution of the linear gap equation (\ref{gap}) with an
adjustable transmission rate between the Pb and the alkali film would be
desirable. Presently we are developing the software for such a solution.

Acknowledgment: The research was supported by NSF Grant No. DMR-0124422.

\newpage

\section{Appendix}

In close vicinity of the transition temperature the superconducting gap
function $\Delta \left( {\bf r}\right) $ is very small and the ''gap
equation'' can be linearized \cite{G38}. 
\begin{eqnarray}
\Delta \left( {\bf r}\right) &=&V\left( {\bf r}\right) \int d^{3}{\bf r}%
^{\prime }\frac{1}{\tau _{T}}\sum_{\omega }H_{\omega }\left( {\bf r,r}%
^{\prime }\right) \Delta \left( {\bf r}^{\prime }\right)  \label{lge} \\
H_{\omega }\left( {\bf r,r}^{\prime }\right) &=&G_{\omega }^{*}\left( {\bf %
r,r}^{\prime }\right) G_{\omega }\left( {\bf r,r}^{\prime }\right) \\
\frac{1}{\tau _{T}} &=&\frac{2\pi k_{B}T}{\hbar }
\end{eqnarray}
Here $\Delta \left( {\bf r}\right) $ is the gap function at the position $%
{\bf r}$, $\omega _{n}=\left( n+1/2\right) /\tau _{T}$ are the Matsubara
frequencies, $N\left( {\bf r}^{\prime }\right) $ is the (BCS)-density of
states for one spin direction, $V\left( {\bf r}\right) $ is the effective
electron-electron interaction at position ${\bf r}$. The function $H_{\omega
}\left( {\bf r,r}^{\prime }\right) $ is the product of the two single
electron Green functions $G_{\omega }\left( {\bf r,r}^{\prime }\right) $ of
a Cooperon. Following de Gennes \cite{D12}, \cite{D32}, Lueders \cite{L14}
and one of the authors \cite{B22} we use a different approach to solve the
gap equation. Since the $G_{w}$ represent the amplitude of an electron
traveling (at finite temperarture) from ${\bf r}^{\prime }$ to ${\bf r}$ the
product $G_{\omega }^{*}\left( {\bf r,r}^{\prime }\right) G_{\omega }\left( 
{\bf r,r}^{\prime }\right) $ describes the pair amplitude of the Cooperon to
travel from ${\bf r}^{\prime }$ to ${\bf r}$. Since the two $G_{w}$ are
conjugate complex to each other and are independent of the spin direction,
the pair amplitude is identical to the probability of a single electron to
travel from ${\bf r}^{\prime }$ to ${\bf r}$. The function $H_{\omega
}\left( {\bf r,r}^{\prime }\right) $ can be expressed by the function $%
F\left( {\bf r,}0{\bf ;r}^{\prime },t^{\prime }\right) $ which gives the
probability of an electron to travel from ${\bf r}^{\prime }$ to ${\bf r}$
during the time interval $\left| t^{\prime }\right| $ (departing at ${\bf r}%
^{\prime }$ at the negative time $t^{\prime }$ and arriving ar ${\bf r}$ at
the time $t=0$) while it experiences an exponential damping of $\exp \left(
2\left| \omega \right| t^{\prime }\right) $. 
\[
H_{\omega }\left( {\bf r,r}^{\prime }\right) =\int_{-\infty }^{0}dt^{\prime
}e^{2\left| \omega \right| t^{\prime }}F\left( {\bf r,}0{\bf ;r}^{\prime
},t^{\prime }\right) N\left( {\bf r}^{\prime }\right) 
\]
This yields the gap equation 
\begin{equation}
\Delta \left( {\bf r}\right) =V\left( {\bf r}\right) \int d^{3}{\bf r}%
^{\prime }\int_{-\infty }^{0}\frac{dt^{\prime }}{\tau _{T}}\sum_{\omega
}e^{2\left| \omega \right| t^{\prime }}F\left( {\bf r,}0{\bf ;r}^{\prime
},t^{\prime }\right) N\left( {\bf r}^{\prime }\right) \Delta \left( {\bf r}%
^{\prime }\right)  \label{gap}
\end{equation}
The function $\sum_{\omega }e^{-2\left| \omega \right| t}$ describes the
exponentially decaying coherence of the Cooperons.

This equation has a very transparent interpretation. From a given position $%
{\bf r}^{\prime }$ and at a given time $t^{\prime }<0$ there are $N\left( 
{\bf r}^{\prime }\right) \Delta \left( {\bf r}^{\prime }\right) $ electrons
(representing the pair amplitude) propagating into all directions of the
metal and experiencing scattering. Along the way their number decays
exponentially as $\sum_{\omega }e^{-2\left| \omega \right| t}$. At a given
time, for example at $t=0$ and at each position ${\bf r}$ one sums the
contribution of all surviving electrons from all $\left( {\bf r}^{\prime
},t^{\prime }\right) $, forming the integral $\int d{\bf r}^{\prime }$ $%
\int_{-\infty }^{0}dt^{\prime }$ and multiplies the result with $V\left( 
{\bf r}\right) $. The result has to reproduce self-consistantly everywhere
the gap function $\Delta \left( {\bf r}\right) $.

If one is dealing with a time dependent gap function the derivation of a
time-dependent Ginzburg-Landau equation from $\left( \ref{gap}\right) $ is
straight forward by replacing the time $0$ by $t$ and $\Delta \left( {\bf r}%
\right) ,\Delta \left( {\bf r}^{\prime }\right) $ by $\Delta \left( {\bf r,}%
t\right) ,\Delta \left( {\bf r}^{\prime },t^{\prime }\right) $ \cite{B22}.

Since we are here not interested in time-dependent gap functions we may
shift the time scale by starting the propagation at ${\bf r}^{\prime }$ at
the time $t=0$ and arriving at ${\bf r}$ at the time $t>0$. Then we have 
\begin{equation}
\Delta \left( {\bf r}\right) =V\left( {\bf r}\right) \int d^{3}{\bf r}%
^{\prime }\int_{0}^{\infty }\frac{dt^{\prime }}{\tau _{T}}\sum_{\omega
}e^{-2\left| \omega \right| t}F\left( {\bf r,}t{\bf ;r}^{\prime },0\right)
N\left( {\bf r}^{\prime }\right) \Delta \left( {\bf r}^{\prime }\right)
\end{equation}

The great advantage of this description is the fact that a major part of
solving the gap equation involves only the dynamics of the conduction
electrons. We demonstrate this first for the Cooper limit of an SN-sandwich.

{\bf Cooper case without barrier}: We assume that the electron-electron
interaction in the normal metal is zero $V_{n}=0$. In this case the gap
function is only non-zero in the superconductor and both ${\bf r}^{\prime }$
and ${\bf r}$ lie in the superconducting film. Since in the Cooper limit $%
\Delta \left( {\bf r}^{\prime }\right) $ is constant and has the value $%
\Delta _{s}$ in the superconducting film one obtains for the gap equation

\[
\Delta _{s}=V_{s}\int_{S}d^{3}{\bf r}^{\prime }\int_{0}^{\infty }\frac{dt}{%
\tau _{T}}\sum_{\omega }e^{-2\left| \omega \right| t}F\left( {\bf r,}t{\bf ;r%
}^{\prime },0\right) N_{s}\Delta _{s} 
\]
The integration $d{\bf r}^{\prime }$ extends only over the superconducting
film. The electron propagation function $F\left( {\bf r,}t{\bf ;r}^{\prime
},0\right) $ describes the probability of ($N\left( {\bf r}^{\prime }\right)
\Delta _{s}$ electrons which start at $\left( {\bf r}^{\prime },0\right) $
(in the superconductor) and arrive at $t$ at the position ${\bf r}$ (in S).
(Although the start and end points of the electron path lie in the
superconductor the path can extend into the normal conductor as well.) Since
the electrons propagate roughly with the Fermi velocity perpendicular to the
film plane the time to move the distance $d_{s} $ or $d_{n}$ is very short
compared with the thermal coherence time $\tau _{T}=\hbar /\left( 2\pi
k_{B}T\right) $. Therefore any electron - independently of where it started
- will be found, after a very short time, in the superconductor with the
probability $p_{s}=\frac{d_{s}N_{s}}{d_{s}N_{s}+d_{n}N_{n}}$. Therefore one
obtains $\int_{S}d^{3}{\bf r}^{\prime }F\left( {\bf r,}t{\bf ;r}^{\prime
},0\right) =\frac{d_{s}N_{s}}{d_{s}N_{s}+d_{n}N_{n}}$ for (almost) any time $%
t$. This yields the equation 
\begin{eqnarray}
\Delta _{s} &=&V_{s}N_{s}\frac{d_{s}N_{s}}{d_{s}N_{s}+d_{n}N_{n}}%
\int_{0}^{\infty }\frac{dt^{\prime }}{\tau _{T}}\sum_{\omega }e^{2\left|
\omega \right| t^{\prime }}\Delta _{s}  \nonumber \\
1 &=&V_{s}N_{s}\frac{d_{s}N_{s}}{d_{s}N_{s}+d_{n}N_{n}}\frac{1}{\tau _{T}}%
\sum \frac{1}{2\left| \omega \right| }  \label{Cooper} \\
\text{with }\frac{1}{\tau _{T}}\sum \frac{1}{2\left| \omega \right| }
&=&\sum_{n=0}^{n_{c}}\frac{1}{n+\frac{1}{2}}  \nonumber
\end{eqnarray}
The equation (\ref{Cooper}) is the well known Cooper condition for the
transition temperature of the sandwich where the attractive interaction is
given by $V_{eff}=V_{s}\frac{d_{s}N_{s}}{d_{s}N_{s}+d_{n}N_{n}}$, resulting
in the transition temperature $T_{c}=1.14\Theta _{D}\exp \left[
-1/N_{s}V_{eff}\right] $. For the initial slope in the Cooper limit one
obtains then 
\[
\frac{d_{s}}{T_{s}}\frac{dT_{c}}{dd_{n}}=-\frac{1}{V_{s}N_{s}}\frac{N_{n}}{%
N_{s}} 
\]
It should be pointed out that the initial slope in the Cooper limit does not
exactly agree with the result by Werthamer, emphasizing that Werthamer's
solution contains approximations.

{\bf Cooper case with barrier: }In the next step we consider the Cooper case
with a barrier between the superconducting and the normal metal film. An
electron in S has a finite transmission rate through the interface which is
proportional to the density of states in the normal conductor $N_{n}$ and
inversely proportional to the thickness of the superconductor $d_{s}$. 
\[
\frac{1}{\tau _{sn}}=\alpha \frac{1}{d_{s}}N_{n} 
\]
We follow the fate of such an electron, denoting the probability to be in S
or N as $n_{s}$ and $n_{n}$, where $\left( n_{s}+n_{n}\right) =1$. Then we
have 
\begin{eqnarray*}
\frac{dn_{s}}{dt} &=&-\alpha n_{s}\frac{1}{d_{s}}N_{n}+\alpha n_{n}\frac{1}{%
d_{n}}N_{s} \\
\frac{dn_{n}}{dt} &=&\alpha n_{s}\frac{1}{d_{s}}N_{n}-\alpha n_{n}\frac{1}{%
d_{n}}N_{s}
\end{eqnarray*}
The solution of this equation is 
\begin{eqnarray}
n_{s} &=&n_{\infty }+n_{\Delta }\exp \left( -\frac{t}{\tau _{r}}\right)
\label{relax} \\
\text{with }n_{\infty } &=&\frac{N_{s}d_{s}}{\left(
N_{n}d_{n}+N_{s}d_{s}\right) }\text{, }n_{\Delta }=\frac{d_{n}N_{n}}{\left(
N_{n}d_{n}+N_{s}d_{s}\right) }  \nonumber \\
\text{ }\frac{1}{\tau _{r}} &=&\alpha \left( \frac{N_{n}}{d_{s}}+\frac{N_{s}%
}{d_{n}}\right) =\frac{1}{\tau _{sn}}+\frac{1}{\tau _{ns}}  \nonumber
\end{eqnarray}
where $\tau _{r}^{-1}$ is the Cooperon relaxation rate. It the sum of the
transmission rates through the interface in both directions. This yields the
following gap equation 
\[
\Delta _{s}=V_{s}N_{s}\int_{0}^{\infty }\frac{dt}{\tau _{T}}\sum_{\omega
}\left( n_{\infty }e^{-2\left| \omega \right| t}+n_{\Delta }e^{-\left(
2\left| \omega \right| +\frac{1}{\tau _{r}}\right) t}\right) \Delta _{s} 
\]
or 
\begin{equation}
1=\frac{V_{s}N_{s}}{d_{s}N_{s}+d_{n}N_{n}}\left( d_{s}N_{s}\sum_{n=0}^{n_{c}}%
\frac{1}{\left( n+\frac{1}{2}\right) }+d_{n}N_{n}\sum_{n=0}^{n_{c}}\frac{1}{%
\left( n+\frac{1}{2}+\frac{1}{2}\frac{\tau _{T}}{\tau _{r}}\right) }\right)
\label{CpBa0}
\end{equation}
This equation shows already qualitatively the effect of the barrier for the
two limiting cases:

\begin{itemize}
\item  weak transmission, $\tau _{T}/\tau _{r}<1$: In this case the sum in
the second term on the right side of equation (\ref{CpBa0}) is almost the
sum as the first sum. One obtains almost the transition temperature of the
pure superconductor. $T_{c}$ is given by the inplicit equation 
\begin{equation}
1=V_{s}N_{s}\sum_{n=0}^{n_{c}}\frac{1}{n+\frac{1}{2}}-V_{s}N_{s}\frac{%
d_{n}N_{n}}{\left( N_{n}d_{n}+N_{s}d_{s}\right) }\left[ \psi \left( \frac{1}{%
2}+\frac{1}{2}\frac{\tau _{T}}{\tau _{r}}\right) -\psi \left( \frac{1}{2}%
\right) \right]  \nonumber
\end{equation}

\item  strong transmission, $\tau _{T}/\tau _{r}>1$: In this case one
obtains 
\[
1=V_{s}N_{s}\frac{d_{s}N_{s}}{\left( N_{n}d_{n}+N_{s}d_{s}\right) }%
\sum_{n=0}^{n_{c}}\frac{1}{n+\frac{1}{2}}+\frac{d_{n}N_{n}}{\left(
N_{n}d_{n}+N_{s}d_{s}\right) }\sum_{n=0}^{n_{c}}\frac{1}{n+\frac{1}{2}+\frac{%
1}{2}\frac{\tau _{T}}{\tau _{r}}} 
\]
where the second term on the right side can be neglected for large $\tau
_{T}/\tau _{r}$ and the remaining part yields just the Cooper transition
temperature without a barrier.
\end{itemize}

We perform the numeric solution of equation (\ref{CpBa0}) for the Pb/Cs
sandwich. For this purpose one has to solve the equation 
\begin{equation}
d_{s}N_{s}\left( \sum_{n=0}^{n_{c}}\frac{1}{\left( n+\frac{1}{2}\right) }-%
\frac{1}{V_{s}N_{s}}\right) +d_{n}N_{n}\left( \sum_{n=0}^{n_{c}}\frac{1}{%
\left( n+\frac{1}{2}+\frac{1}{2}\frac{\tau _{T}}{\tau _{r}}\right) }-\frac{1%
}{V_{s}N_{s}}\right) =0  \label{GE_3a}
\end{equation}
with 
\begin{eqnarray}
\text{ }n_{c} &=&\frac{\Theta _{D}}{2\pi T_{c}}\text{, }n_{c0}=\frac{\Theta
_{D}}{2\pi T_{s}}  \label{GE_3b} \\
\rho &=&\frac{\tau _{T}}{\tau _{r}}=\frac{\tau _{T}}{\tau _{sn}}\left( 1+%
\frac{d_{s}N_{s}}{d_{n}N_{n}}\right)  \nonumber \\
\frac{1}{V_{s}N_{s}} &=&\sum_{n=0}^{n_{c0}}\frac{1}{\left( n+\frac{1}{2}%
\right) }  \nonumber
\end{eqnarray}
The resulting $T_{c}$-values are plotted in Fig.2a and Fig.3 as full curves.
For the transmission rate we used the value $1/\tau _{sn}=.035/\tau
_{T_{s}}=.035\ast 2\pi k_{B}T_{s}/\hbar $ $=2.0\times 10^{11}s^{-1}$ or $%
\tau _{sn}=5\times 10^{-12}s$. ($1/\tau _{sn}$ is transmission rate for an
electron in the Pb film to escape into the Cs film). It is much longer than
the superconducting time constant $\tau _{T_{s}}=\hbar /\left( 2\pi
k_{B}T_{s}\right) $ $=1.7\times 10^{-13}s$ and the ballistic flight time of
the Pb electrons with (normalized) Fermi velocity (in z-direction): $\tau
_{b}=2d/v_{F}^{\ast }=4\times 10^{-14}s$.

Similar evaluations are performed for the Pb/Rb and Pb/K sandwiches. The
transmission rate ratios $\tau _{T_{s}}/\tau _{sn}$ are included in table I.
For the Pb/Na sandwich the agreement was not sufficient to include it in the
Fig.2d. All the theoretical curves for the Cooper case with barrier yield a
too large initial slope. For small thicknesses of the normal conductor the
relaxation rate becomes arbitrarily large according to relation (\ref{relax}%
). This is unphysical, however, because the electron needs at least the
ballistic flight time $\tau _{b}$ to cross over from the superconductor into
the normal conductor. Therefore this range has to be excluded from the
evaluation. Going from Pb/Cs to Pb/Na the alkali thickness range in which
the barrier approach is not appropriate increases.

Ashida et al. \cite{A57} calculated the transition temperature of
SN-sandwiches with barriers in-between.. They characterize the strength of
the barrier by the coefficient of reflectivity $R$ at the interface. (In the
absence of a barrier the value of $R$ is not zero but given by density of
states ratio). They find in the limit $\left( R-1\right) <<1$ for the
transition temperature (in our notation) 
\[
\frac{\Delta T_{c}}{T_{s}}=\frac{\pi ^{2}}{16}\frac{\hbar }{2\pi k_{B}T_{c}}%
\frac{\left( 1-R\right) v_{s}}{d_{s}} 
\]

This limit $\left( R-1\right) <<1$ corresponds to in our notation to the
limit $1/\tau _{sn}<<1/\tau _{T}$. If we follow Ashida et al. and expand \ref
{GE_3a} in terms of the pair breaking parameter $\rho $ we obtain

\[
\sum_{n=0}^{n_{c}}\frac{1}{\left( n+\frac{1}{2}+\frac{1}{2}\rho \right) }%
\thickapprox \sum_{n=0}^{n_{c}}\frac{1}{n+\frac{1}{2}}-\frac{\rho }{2}*\frac{%
1}{2}\pi ^{2} 
\]
\begin{equation}
\left( \frac{d_{n}N_{n}}{d_{s}N_{s}}\right) ^{-1}=-1+\frac{\frac{\rho }{2}*%
\frac{1}{2}\pi ^{2}}{\left( \sum_{n=0}^{n_{c}}\frac{1}{\left( n+\frac{1}{2}%
\right) }-\frac{1}{V_{s}N_{s}}\right) }
\end{equation}
which yields 
\[
\ln \left( \frac{T_{s}}{T_{n}}\right) =\frac{\pi ^{2}}{4}\frac{\tau _{T}}{%
\tau _{sn}} 
\]
or 
\[
\frac{\Delta T_{c}}{T_{s}}=\frac{\pi ^{2}}{4}\frac{\hbar }{2\pi k_{B}T_{c}}%
\frac{1}{\tau _{sn}} 
\]
The calculation of $\overline{1/\tau _{sn}}$ in terms of $R$ is straight
forward 
\[
\overline{\frac{1}{\tau _{sn}}}=\frac{\int_{0}^{1}\frac{v_{F}\cos \theta }{%
2d_{s}}\left( 1-R\right) d\left( \cos \theta \right) }{\int_{0}^{1}d\left(
\cos \theta \right) }=\frac{v_{F}\left( 1-R\right) }{4d_{s}} 
\]
This expansion yields a perfect agreement between the results of Ashida et
al. and ours. However, the result is somewhat surprising since the $T_{c}$
does not depend on the thickness of the normal conductor. Our numerical
evaluation shows that this is not quite correct. The reason is the following:

The value of $\rho =\frac{\tau _{T}}{\tau _{r}}=\frac{\tau _{T}}{\tau _{sn}}%
\left( 1+\frac{d_{s}N_{s}}{d_{n}N_{n}}\right) $ diverges for small thickness
of the normal conductor, always reaching the limit of strong transmission
for $d_{n}\rightarrow 0$. That means that at small thickness of the normal
conductor one expects the full Cooper reduction of the transition
temperature. This means in other words that even for $\left( R-1\right) <<1$
the expansion by Ashida et al. has to exclude the range of very small normal
conductor thickness. When $d_{n}$ increases the transmission into the normal
conductor becomes less effective, the superconducting film behaves more
isolated and the $T_{c}$-reduction is much less effective. This is the
experimental observation, in particular for the Pb/Cs sandwich.

\newpage

\section{Figures}

\FRAME{dtbpF}{4.4348in}{3.2958in}{0pt}{}{}{f_arb77_01.eps}{\special{language
"Scientific Word";type "GRAPHIC";maintain-aspect-ratio TRUE;display
"USEDEF";valid_file "F";width 4.4348in;height 3.2958in;depth
0pt;original-width 4.2082in;original-height 3.1202in;cropleft "0";croptop
"1";cropright "1";cropbottom "0";filename '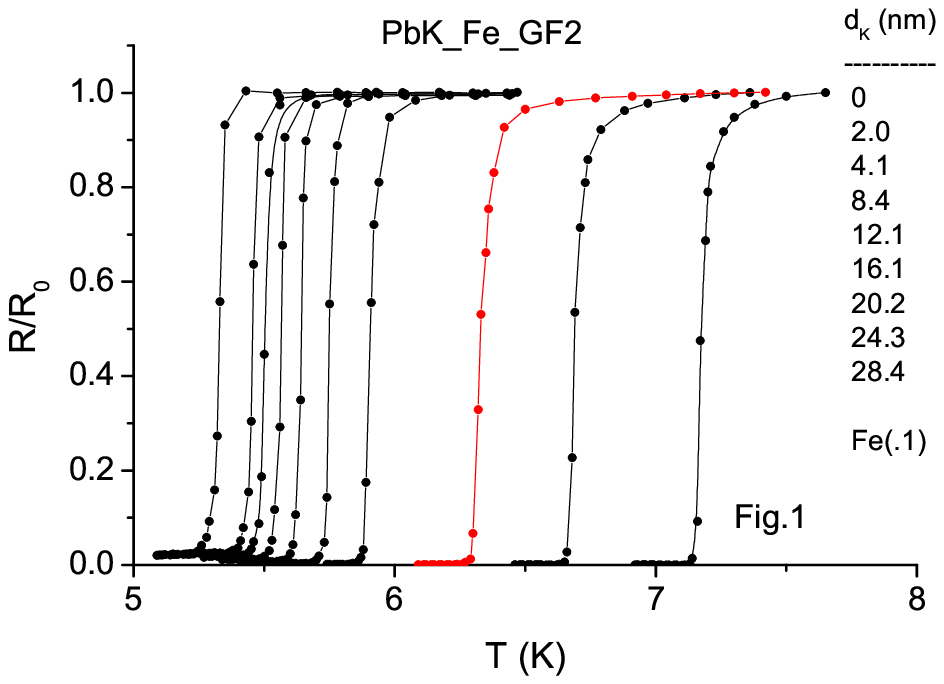';file-properties
"XNPEU";}}Fig.1: The superconducting transition curves for Pb with different
coverages of K.

\[
\begin{array}{ll}
\FRAME{itbpF}{2.4647in}{2.0202in}{0pt}{}{}{f_arb77_02a.eps}{\special%
{language "Scientific Word";type "GRAPHIC";maintain-aspect-ratio
TRUE;display "USEDEF";valid_file "F";width 2.4647in;height 2.0202in;depth
0pt;original-width 3.8597in;original-height 3.1609in;cropleft "0";croptop
"1";cropright "1";cropbottom "0";filename '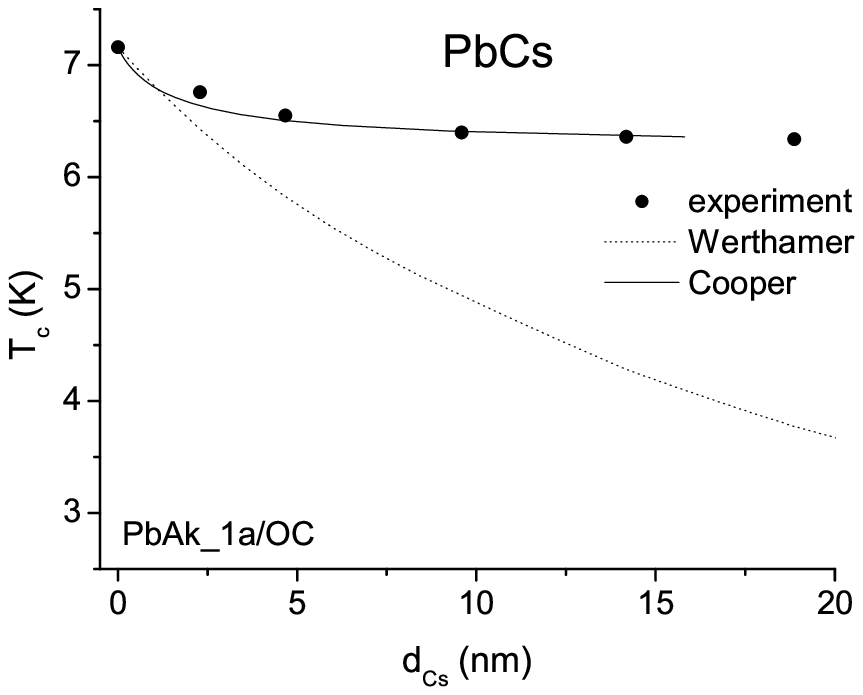';file-properties
"XNPEU";}} & \FRAME{itbpF}{2.4388in}{1.9995in}{0pt}{}{}{f_arb77_02b.eps}{%
\special{language "Scientific Word";type "GRAPHIC";maintain-aspect-ratio
TRUE;display "USEDEF";valid_file "F";width 2.4388in;height 1.9995in;depth
0pt;original-width 3.8597in;original-height 3.1609in;cropleft "0";croptop
"1";cropright "1";cropbottom "0";filename '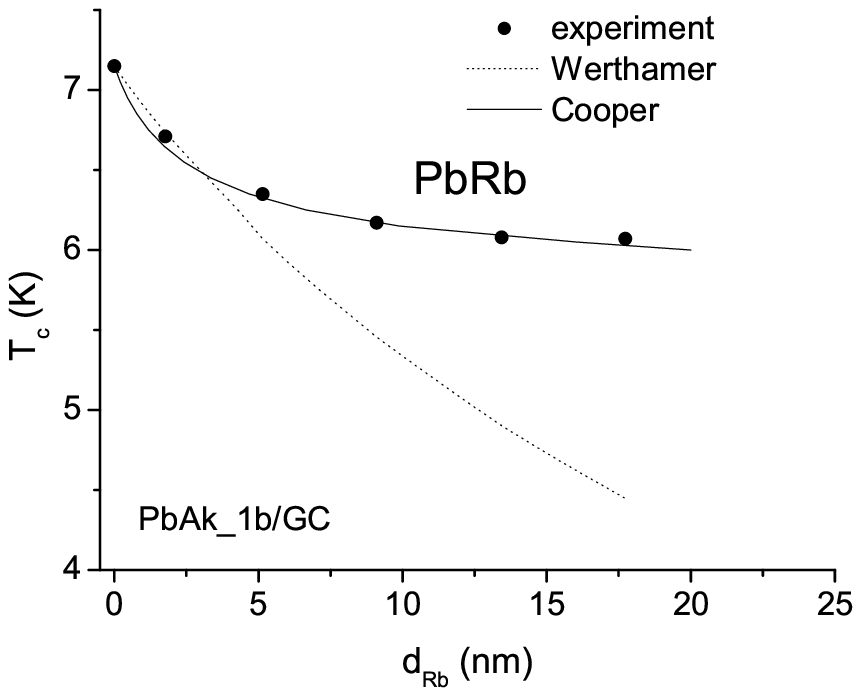';file-properties
"XNPEU";}} \\ 
\FRAME{itbpF}{2.4993in}{2.0496in}{0pt}{}{}{f_arb77_02c.eps}{\special%
{language "Scientific Word";type "GRAPHIC";maintain-aspect-ratio
TRUE;display "USEDEF";valid_file "F";width 2.4993in;height 2.0496in;depth
0pt;original-width 3.8597in;original-height 3.1609in;cropleft "0";croptop
"1";cropright "1";cropbottom "0";filename '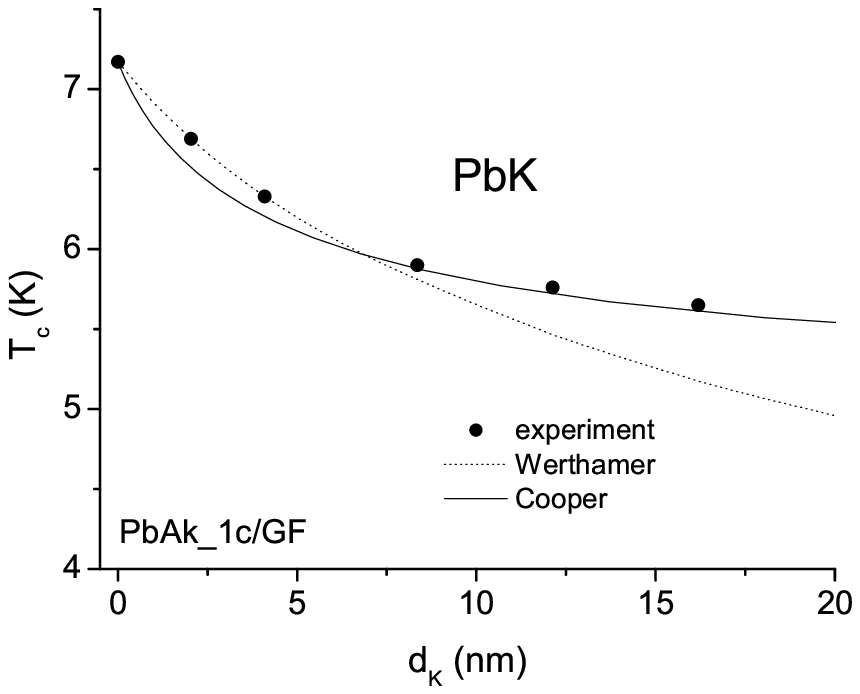';file-properties
"XNPEU";}} & \FRAME{itbpF}{2.4405in}{2.009in}{0pt}{}{}{f_arb77_02d.eps}{%
\special{language "Scientific Word";type "GRAPHIC";maintain-aspect-ratio
TRUE;display "USEDEF";valid_file "F";width 2.4405in;height 2.009in;depth
0pt;original-width 3.845in;original-height 3.1609in;cropleft "0";croptop
"1";cropright "1";cropbottom "0";filename '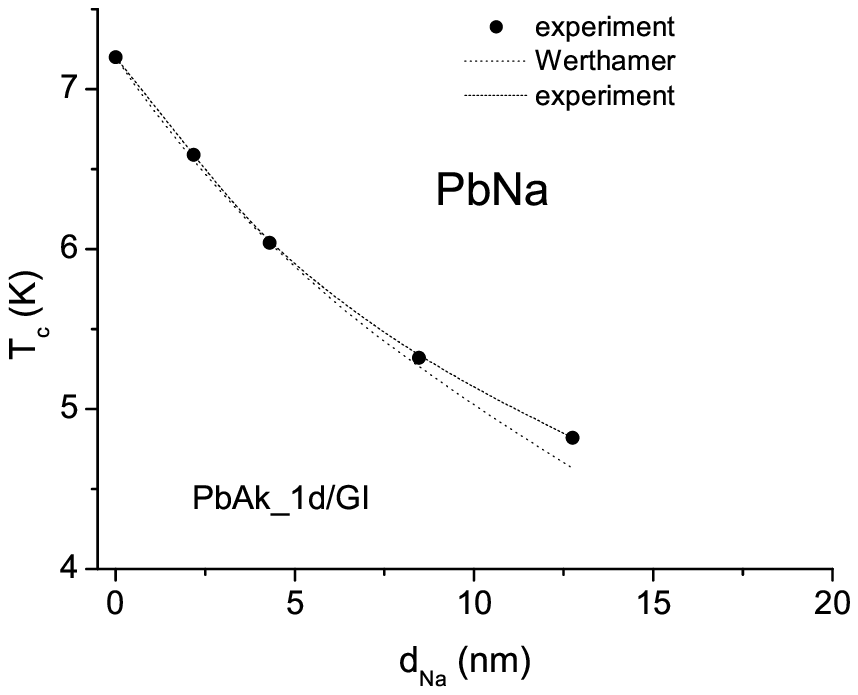';file-properties
"XNPEU";}}
\end{array}
\]
Fig.2a-d: The transition temperatures of Pb/Ak multilayers as a function of
the alkali thickness. The dotted curves are calculated with the theory by
Werthamer. For Pb/Cs, Pb/Rb and Pb/K the full curves are calculated within
the Cooper model with a barrier between the Pb and the alkali film. For
Pb/Na the full curve is a guide to the eye.\FRAME{dtbpF}{3.3996in}{2.7363in}{%
0pt}{}{}{f_arb77_03.eps}{\special{language "Scientific Word";type
"GRAPHIC";maintain-aspect-ratio TRUE;display "USEDEF";valid_file "F";width
3.3996in;height 2.7363in;depth 0pt;original-width 4.2627in;original-height
3.4264in;cropleft "0";croptop "1";cropright "1";cropbottom "0";filename
'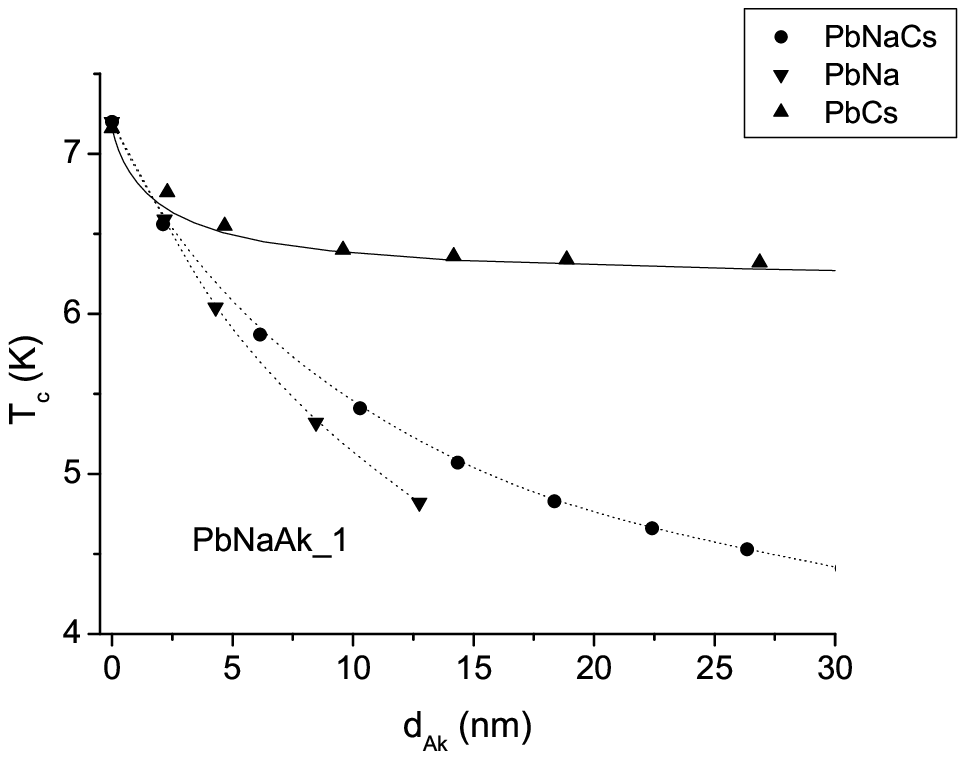';file-properties "XNPEU";}}Fig.3: The transition
temperatures of Pb/Na/Cs multilayers as a function of the alkali thickness
(full circles). The transition temperatures of the Pb/Cs sandwiches (up
triangles) and the Pb/Na sandwiches (down triangles) are shown for
comparison. The full curve for the PbCs is calculated in the Cooper model
with a barrier between the Pb and Cs films.

\end{document}